%% file: omega_eps2005.tex
\documentclass[aps,prl,preprint,tightenlines,showpacs,a4paper,floats,superscriptaddress,twocolumn]{revtex4}
\usepackage{epsfig}
{
\begin{document}
\preprint{\vbox{ \hbox{   }
                 \hbox{BELLE-CONF-0526}
		 \hbox{LP2005-161}
		 \hbox{EPS2005-499}
}}
\begin{flushleft}
\includegraphics[width=1.5in]{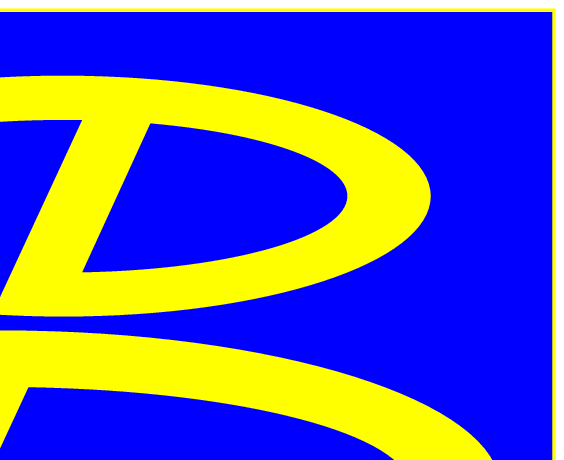}
\end{flushleft}    % BELLE-logo
\title{
\quad\\[0.5cm] %\Large
\boldmath
Improved Measurements of Branching Fractions and $CP$ Asymmetries for $B \to \omega h$ Decays.}
%
\input{author-conf2005.tex}
\date{\today}
\begin{abstract}
We report improved measurements of $B$ to pseudoscalar-vector decays containing 
an $\omega$ meson in the final state. 
These results are obtained from a data sample that contains 386 million $B\bar{B}$ 
pairs collected at the $\Upsilon(4S)$ resonance, with the Belle detector at the KEKB 
asymmetric energy $e^+ e^-$ collider. We measure the branching fractions 
${\mathcal B}(B^\pm\to \omega K^\pm) = (8.1 \pm 0.6 \pm 0.5)\times 10^{-6}$, 
${\mathcal B}(B^\pm \to \omega \pi^\pm) = (7.0 \pm 0.6 \pm 0.5)\times 10^{-6}$, and
${\mathcal B}(B^0 \to \omega K^0) = (3.9 \pm 0.7 \pm 0.4 )\times 10^{-6}.$ 
We also set the 90\% confidence level upper limit ${\mathcal B}(B^0 \to \omega \pi^0) < 1.5\times 10^{-6}.$ 
In addition, we obtain the partial rate 
asymmetries ${\mathcal A}_{CP} = 0.05 \pm 0.08 \pm 0.01$ for $B^\pm \to \omega K^\pm$
and ${\mathcal A}_{CP} = -0.03 \pm 0.09 \pm 0.02$ for $B^\pm\to \omega \pi^\pm.$
\end{abstract}
\maketitle
\tighten
\normalsize
\setcounter{footnote}{0}
\normalsize
Charmless hadronic $B$ decays provide a rich ground to understand the dynamics of $B$ meson decays and the origin of $CP$ violation.
Two-body $B$ decays with a vector meson $\omega$ and a pseudoscalar particle $h$ 
($h$ can be kaon or pion) proceed through $b\to s$ penguin diagrams and 
$b\to u$ diagrams. Theoretical expectations from QCD factorization 
approaches suggest that the branching fractions of charged $B$ decays
are between $3$ and $8\ \times 10^{-6}$ and that the $\omega K^+$ branching fraction is smaller than
the $\omega \pi^+$ \cite{QCDF,PQCD}. The decay $B^0\to \omega \pi^0$ is color
suppressed while the neutral $B$ decay to $\omega K^0$ is expected to have 
slightly smaller branching fraction than the charged $B$ decays.
Experimentally, clear signals have been observed in 
$B^\pm \to \omega K^\pm, B^\pm \to \omega \pi^\pm$ and $B^0\to \omega K^0$ with
similar branching fractions \cite{chwang, previous, babar}. 
However, experimental measurements
are not yet precise enough to tell which mode has the larger branching fraction. 
Furthermore, the branching fraction of $B^0\to\omega K^0$ measured by the Belle 
collaboration is smaller than the BaBar result. Therefore, it is necessary to update the results with more data.\par   

In this paper, we report improved measurements of branching fractions and 
partial rate asymmetries for $B \to \omega h$ decays, where $h$ can 
be a kaon or pion. The partial rate asymmetry (${\mathcal A}_{CP}$) is measured for the charged $B$ decays and defined to be 
\begin{eqnarray}  
{\mathcal A}_{CP} \equiv \frac{\Gamma(B^- \to f^-)-\Gamma(B^+ \to f^+)}{\Gamma(B^- \to f^-)+\Gamma(B^+ \to f^+)},
\end{eqnarray}
where $\Gamma(B^- \to f^-)$ is the $B^-$ decay rate and $\Gamma(B^+ \to f^+)$ denotes that of the charge conjugate mode.   
These measurements are based on a data sample of 386 million $B\bar{B}$ pairs, collected with the Belle detector at the KEKB~\cite{KEKB} 
asymmetric-energy $e^+e^-$ collider. The statistics are a factor of 5 larger than our previous published results.\par

The Belle detector is a large-solid-angle magnetic spectrometer that 
consists of a silicon vertex detector (SVD), a 50-layer central drift chamber 
(CDC), an array of aerogel threshold \v{C}erenkov counters (ACC), a barrel-like 
arrangement of time-of-flight scintillation counters (TOF), and an 
electromagnetic calorimeter comprised of CsI(Tl) crystals (ECL) located inside 
a super-conducting solenoid coil that provides a 1.5~$\rm T$ magnetic field. An iron 
flux-return located outside of the coil is instrumented to detect $K_L^0$ 
mesons and to identify muons (KLM).  The detector is described in detail 
elsewhere~\cite{Belle}.
In August 2003, the three-layer SVD was replaced by a four-layer radiation tolerant device. The data sample for this analysis consists of 
140 $\rm {fb}^{-1}$ of data with the old SVD (Set I) and 217 $\rm {fb}^{-1}$ with the new one (Set II).\par

Hadronic events are selected using criteria based on the charged track 
multiplicity and total visible energy with an efficiency 
greater than 99\% for generic $B\bar{B}$ events. All primary charged tracks must satisfy quality requirements based on their 
impact parameters relative to the run-dependent interaction point (IP). For  
tracks from the candidate $B$ mesons, their deviations from the IP 
position are required to be within $\pm 0.1$ cm in the transverse direction and $\pm 3.0 
\rm {cm}$ in the longitudinal direction. Charged particle identification is performed using the $K$-$\pi$ likelihood ratio,  
${\mathcal R}_{K} = {\mathcal L}_{K}$/(${\mathcal L}_{\pi}+{\mathcal L}_{K}$), where ${\mathcal L}_{K}$ 
(${\mathcal L}_{\pi}$) is the likelihood of a charged particle to be a kaon (pion) based on the information coming from ACC, 
TOF and CDC. Charged tracks from $B$ meson candidates with ${\mathcal R}_{K} > 0.6$ are identified as kaons 
and ${\mathcal R}_{K} < 0.4$ as pions. Kaons produced in the $B^\pm \to \omega K^\pm$ decays are selected by these criteria
with efficiencies $85\%$, while the corresponding rates of Kaons that are misidentified as pions are $11\%$; on the other hand,
pions from $B^\pm \to \omega \pi^\pm$ are selected with an efficiency of $90\%$ and a kaon misidentification rate of $9\%$. 
Candidate $\pi^0$ mesons are reconstructed from pairs of photons with invariant mass in the range 
$0.1178\ \rm {GeV/}c^2 <$ $M_{\gamma\gamma}$ $< 0.1502\ \rm {GeV/}c^2$. Candidate $K^0_S$ mesons are reconstructed using 
pairs of oppositely charged particles that have an invariant mass in the range 
$0.482\ \rm {GeV/}c^2 <$ $M_{\pi^+\pi^-}$ $< 0.514\ \rm {GeV/}c^2$. The vertex of the $K^0_S$ candidate is required to be 
well reconstructed and displaced from the interaction point, and the $K^0_S$ momentum direction must be consistent with the $K^0_S$ 
flight direction. Candidate $\omega \to \pi^+ \pi^- \pi^0$ decays are reconstructed from charged tracks with 
${\mathcal R}_{K} < 0.9$, where the pions from the $\omega$ decay are selected with $96\%$ efficiency, 
and the $\pi^0$s with center-of-mass frame momentum are greater than $0.35\ \rm {GeV/}c^2$.  
Candidate $\omega$ mesons are required to have invariant masses within $\pm 30\ \rm {MeV/}c^2$ of the nominal value.\par

$B$ meson candidates are formed by combining an $\omega$ meson with either a kaon ($K^\pm$, $K^0_S$) or a pion ($\pi^\pm$, $\pi^0$).
Two kinematic variables are used to select $B$ candidates: the beam constrained mass 
$M_\mathrm{bc}$ = $\sqrt{(E^{\rm CM}_{\rm beam})^2-|P^{\rm CM}_B|^2}$  and the energy difference 
$\Delta E = E^{\rm CM}_B - E^{\rm CM}_{\rm beam}$, where $E^{\rm CM}_{\rm beam}$ is the beam energy 
in the $\Upsilon(4S)$ rest frame, and $P^{\rm CM}_B$, $E^{\rm CM}_B$ are the momentum and energy of the $B$ candidate in the 
$\Upsilon(4S)$ rest frame. Candidates with $M_\mathrm{bc} > 5.2\ \rm {GeV/}c^2$ and $|\Delta E|<0.25\ \rm {GeV}$ 
($|\Delta E|<0.30\ \rm {GeV}$ for $B^0 \to \omega \pi^0$) are selected for further analysis.\par

The dominant background comes from quark-antiquark continuum events ($e^+e^- \to q\bar{q}$, $q = u$,$d$,$s$,$c$). 
The continuum background is characterized by a jet-like structure while the 
$B\bar{B}$ events have a more spherical distribution.
Several event-shape variables are employed to suppress the continuum. The 
thrust angle $\theta_T$ is defined as the angle between the 
thrust axis \cite{thrust} of the $B$ candidate daughter particles and that of the rest of 
the particles in an event. Signal events are uniformly distributed in 
$\cos\theta_T$, while continuum events are sharply peaked near $\cos\theta_T = \pm 1.0$. 
Events with $|\cos\theta_T| < 0.9$ are selected. 
A Fisher discriminant is formed by combining    
a set of modified Fox-Wolfram moments \cite{SFW} with the variable 
$S_{\perp}$, defined as the scalar sum of the transverse 
momentum of all particles outside a $45^\circ$ cone around the thrust axis of the
$B$ candidate divided by the scalar sum of their momenta. 
Further variables that have been found to separate signal from continuum background include: 
the cosine of the angle between the flight 
direction of the $B$ candidate and the beam direction ($\cos\theta_B$), 
the distance along the beam direction between the $B$ vertex and the vertex of
the remaining particles of the event ($\Delta z$) and the cosine of the helicity 
angle, defined as the angle between the $B$ flight direction in the $\omega$
rest frame and the normal direction of the plane spanned by the three daughter
pions of the $\omega$. The probability density functions (PDFs) for these 
three variables and the Fisher discriminant are obtained using events in signal 
Monte Carlo simulation and data with $M_\mathrm{bc}< 5.26\ \rm {GeV/}c^2$ for signal and
$q \bar{q}$ background, respectively.
These variables are combined to form a likelihood ratio 
$\mathcal{R} = \frac{{\mathcal L}_{S}}{{\mathcal L}_{S}+{\mathcal L}_{q\bar{q}}}$, where
${\mathcal L}_{S (q\bar{q})}$ is the product of signal ($q \bar{q}$)
probability densities.\par

Additional background discrimination is provided by the quality of
the $B$ flavor tagging of the accompanying $B$ meson.
We use the standard Belle $B$ tagging package \cite{TaggingNIM},
which gives two outputs:
a discrete variable ($q$) indicating the $B$ flavor and a dilution factor ($r$)
ranging from zero for no flavor information to unity for unambiguous
flavor assignment. We divide the data into six $r$ regions.
Continuum suppression is achieved by applying a mode dependent requirement on
${\mathcal R}$ for events in each $r$ region based on
${\mathcal N}_s/\sqrt{{\mathcal N}_s+{\mathcal N}_{q\bar{q}}}$,
where ${\mathcal N}_s$ is the expected number of signal events estimated from simulation and
${\mathcal N}_{q\bar{q}}$ denotes the number of background events estimated from data. 
This ${\mathcal R}$ requirement retains $74$\%, $68$\%, $74$\%, and $57$\% of the signal while 
rejecting $91$\%, $94$\%, $91$\%, and $95$\% of the continuum background for the $\omega K^\pm$, 
$\omega \pi^\pm$, $\omega K^0$, and $\omega \pi^0$ modes, respectively.\par 

Simulation studies indicate small backgrounds from generic $b\to c$ transitions in the
charged $B$ modes; they are found to be negligible for the neutral modes. Moreover,
two other backgrounds need to be considered: 
the $\omega K^+ \leftrightarrow \omega \pi^+$ reflection, due to
$K^+ \leftrightarrow \pi^+$
misidentification, and the feed-down from charmless $B$ decays, predominantly
$B\to \omega K^*(892)$ and  $B\to \omega\rho(770)$.
We include these three components in the fit used to extract the signal.\par

The signal yields and partial rate asymmetries are obtained  
using an extended unbinned maximum likelihood fit on 
$M_\mathrm{bc}$ and $\Delta E$.
The likelihood is defined as 
\begin{eqnarray}
\mathcal{L} & = & \rm e^{-\sum_{j} {\mathcal N}_{j}}
\times \prod_{i} (\sum_{j} {\mathcal N}_{j} \mathcal{P}_{j}) \;\;\; \\
\mathcal{P}_{j} & = &\frac{1}{2}[1- Q_{i} \cdot {\mathcal A}_{CP}{}_{j}]
 P_{j}(M_{{\mathrm bc}i}, \Delta E_{i})
\end{eqnarray}
where $i$ is the identifier of the $i$-th event, $P(M_{\mathrm bc}, \Delta E)$ 
is the PDF of $M_{\mathrm bc}$ and $\Delta E$, $Q$ indicates the $B$ meson charge,
$B^+(Q=+1)$ or $B^- (Q=-1)$, ${\mathcal N}_{j}$ is the number of events for the
category $j$, which corresponds to either signal, $q\bar{q}$ continuum,  
a reflection due to $K$-$\pi$ misidentification, or
$B\bar{B}$ backgrounds, inclusive of $b\to c$ background and other charmless $B$ decays. 
For the neutral $B$ mode, ${\mathcal P}_{j}$ in Eq.(2) is simply
$P_{j}(M_{{\mathrm bc}i}, \Delta E_{i})$ and there is no reflection component and
$b\to c$ background.\par
 
The signal distribution in $M_\mathrm{bc}$ is parameterized by a Gaussian function 
centered near the mass of the $B$ meson, while the Crystal-Ball line shape 
\cite{crystal} is used to model the $\Delta E$ distribution. The product of 
the two functions is used as the signal PDF, calibrated with large control samples of 
$B\to D\pi, \overline{D}^0\to K^+\pi^-\pi^0$ and 
$B\to \eta^\prime (\eta \pi^+\pi^-) K$ decays.
The continuum PDF is the product of a first order polynomial
for $\Delta E$ and an ARGUS function \cite{argus} for $M_\mathrm{bc}$. 
Other background PDFs are modelled by a smoothed two-dimensional $M_\mathrm{bc}-\Delta E$ 
function obtained from the MC simulation. The yields of the 
$B\bar{B}$ background mentioned in the previous paragraph are fixed with ${\mathcal A}_{CP}= 0$ 
in the fit based on the MC simulation except for the reflection component. The ${\mathcal A}_{CP}$ and 
the normalizations of the reflection components are fixed to expectations
based on the $B^+\to \omega K^+$ and $B^+\to \omega\pi^+$ partial rate
asymmetries and branching fractions, as well as
$K^+\leftrightarrow \pi^+$ fake rates.
The reflection yield and ${\mathcal A}_{CP}$ are
first input with the assumed values and are then
recalculated according to our measured results.\par

Table~\ref{eps05-1} and Table~\ref{eps05-2} show the measured branching fractions and 
${\mathcal A}_{CP}$ as well as other quantities associated with the measurements. 
Branching fractions are computed as the sum of the yields divided by the corresponding efficiencies
in each dataset, divided by the total number of $B$ mesons. The reconstruction efficiency for each mode 
is defined as the fraction of the signal yield remaining after all selection criteria, where the yield 
is determined by performing the unbinned maximum likelihood fit on the MC sample. We define the significance 
as $\sqrt{-2{\rm ln}({\mathcal L}_{0}/{\mathcal L}_{\rm max})}$, where ${\mathcal L}_{\rm max}$ is the maximum likelihood 
from the fit when the signal branching fraction is floated, and ${\mathcal L}_{0}$ is the likelihood obtained 
when the signal branching fraction is set to zero. Systematic uncertainties are included by repeating the fit with all parameters 
fixed at $\pm 1 \sigma$ from their central value and choosing the set of parameters that gives the smallest significance.\par

For the branching fraction measurement, the main sources of systematic uncertainties are: 
charged tracking efficiency ($1.0\%$ per track); neutral pion detection ($3.5\%$ for Set I and $4.0\%$ for Set II); 
$K^0_{S}$ reconstruction ($4.5\%$ for Set I and $4.0\%$ for Set II); $\omega$ mass resolution ($0.2\%$ for Set I and $0.6\%$ for Set II); 
requirement on ${\mathcal R}$ (about $3.0\%$). Systematic uncertainties due to fit parameters are evaluated by varying each of the 
fixed parameters by $\pm 1\sigma$, and by adding in quadrature the resulting deviations from the central value. The total systematic
uncertainty is the quadratic sum of all the above contributions.\par       
\begin{table}[htb]
\begin{center}
\caption{Signal yields (${\mathcal N}_s$), efficiencies  ($\epsilon_{tot}$), fitted branching fraction (${\mathcal B}$), 
 partial rate asymmetry (${\mathcal A}_{CP}$), and fit significance including systematic error ($\Sigma$) for the $B^\pm \to \omega K^\pm$
 and $B^\pm \to \omega \pi^\pm$ modes.}
\begin{tabular}{|l c c|}\hline\hline
Mode & $\omega K^\pm$ & $\omega \pi^\pm$\\\hline    
${\mathcal N}_s$ ($\rm {Set I}$)  & $90^{+12}_{-11}$ & $84^{+12}_{-11}$\\                         
${\mathcal N}_s$ ($\rm {Set II}$)  & $170 \pm 16$ & $145 \pm 16$\\
$\epsilon_{tot}$ ($\rm {Set I}$)  & $8.6 \pm 0.1$\ \% & $9.1 \pm 0.1$\ \%\\
$\epsilon_{tot}$ ($\rm {Set II}$)  & $8.5 \pm 0.1$\ \% & $8.6 \pm 0.1$\ \%\\
${\mathcal B}$ ($\times 10^{-6}$) & $8.1\pm 0.6 \pm 0.5$ & $7.0\pm0.6 \pm 0.5$\\
${\mathcal A}_{CP}$ & $0.05\pm0.08 \pm0.01$ & $-0.03 \pm0.09 \pm0.02$\\
$\Sigma$ & $20.0 \sigma$ & $17.3 \sigma$\\\hline\hline
\end{tabular}
\label{eps05-1}
\end{center}
\end{table}
\begin{table}[htb]
\begin{center}
\caption{Signal yields (${\mathcal N}_s$), efficiencies ($\epsilon_{tot}$), fitted branching fraction (${\mathcal B}$), 
 upper limit on the branching fraction (U.L.) including systematic error, and fit significance including 
 systematic error ($\Sigma$) for the $B^0 \to \omega K^0$ and $B^0 \to \omega \pi^0$ modes.}
\begin{tabular}{|l c c|}\hline\hline
Mode & $\omega K^0$ & $\omega \pi^0$\\\hline    
${\mathcal N}_s$ ($\rm {Set I}$)  & $16^{+5.0}_{-4.4}$ & $-$\\                         
${\mathcal N}_s$ ($\rm {Set II}$)  & $14^{+5.1}_{-4.4}$ & $2.8^{+4.7}_{-4.4}$\\
$\epsilon_{tot}$ ($\rm {Set I}$)  & $2.7 \pm 0.1$\ \% & $4.7 \pm 0.1$\ \%\\
$\epsilon_{tot}$ ($\rm {Set II}$)  & $2.8 \pm 0.1$\ \% & $3.7 \pm 0.1$\ \%\\
${\mathcal B}$ ($\times 10^{-6}$) & $3.9\pm0.7\pm0.4$ & $0.8^{+0.5}_{-0.4}\pm0.1$\\
U.L. ($\times 10^{-6}$) & $-$ & $1.5$\\
$\Sigma$ & $9.3 \sigma$ & $2.2 \sigma$\\\hline\hline
\end{tabular}
\label{eps05-2}
\end{center}
\end{table}
In summary, we update the measurements of the branching fractions of the exclusive two-body charmless hadronic $B$ decays 
with an $\omega$ meson in the final state. We measure the following branching fractions: 
${\mathcal B}(B^\pm\to\omega K^\pm)=(8.1\pm0.6\pm0.5)\times10^{-6}$, 
${\mathcal B}(B^\pm\to\omega \pi^\pm)=(7.0\pm0.6\pm0.5)\times10^{-6}$, and ${\mathcal B}(B^0\to\omega K^0)=(3.9\pm0.7\pm0.4)\times10^{-6}$. 
We also set the following upper limit: ${\mathcal B}(B^0\to\omega \pi^0)<1.5\times10^{-6}$ at the 90\% confidence level.
Finally, we measure the partial rate asymmetries in $B^\pm \to \omega K^\pm$ and $B^\pm \to \omega \pi^\pm$:
${\mathcal A}_{CP}= 0.05 \pm 0.08 \pm 0.01$ for $\omega K^\pm$, and ${\mathcal A}_{CP} =  -0.03 \pm 0.09 \pm 0.02$ for 
$\omega \pi^\pm$; no evidences of direct CP violation is found in either mode so far.\par
\begin{figure}
\includegraphics[width=0.52\textwidth]{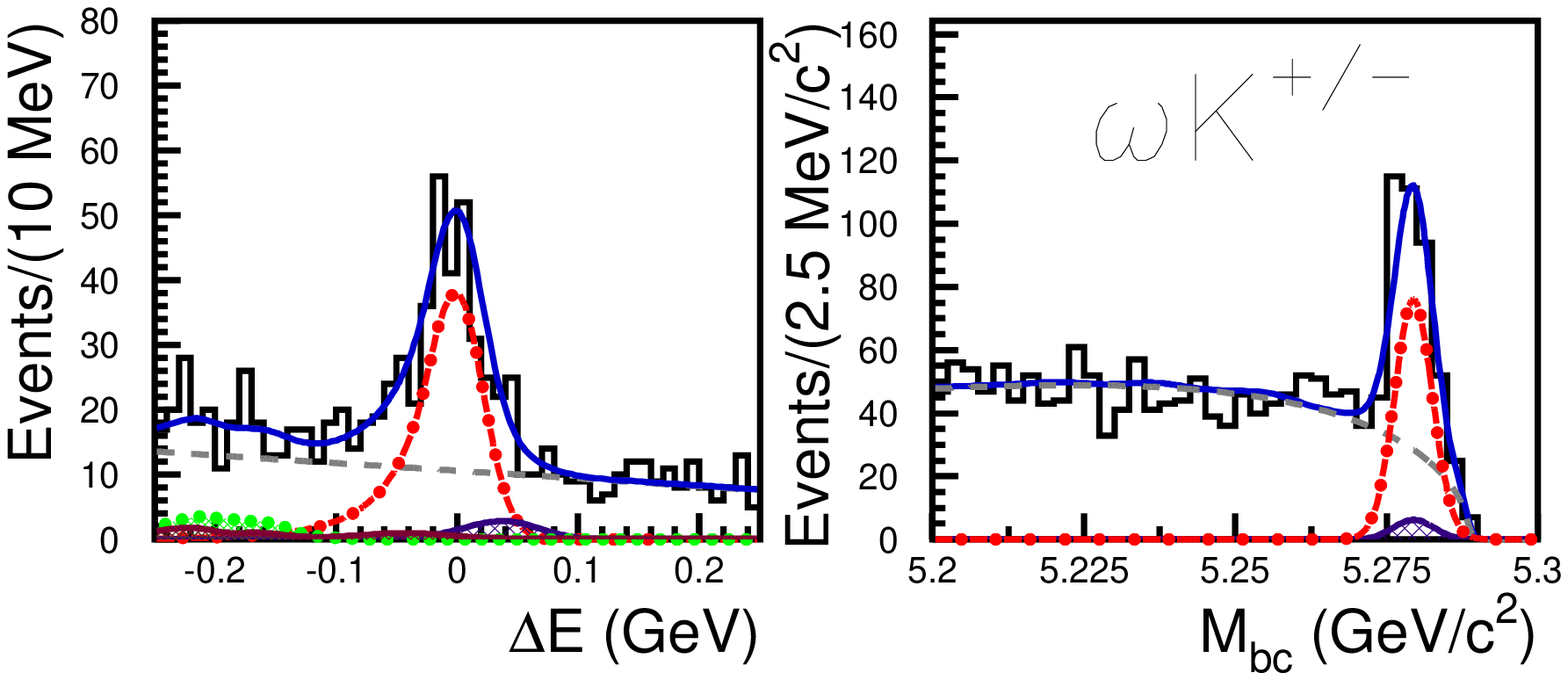}
\includegraphics[width=0.52\textwidth]{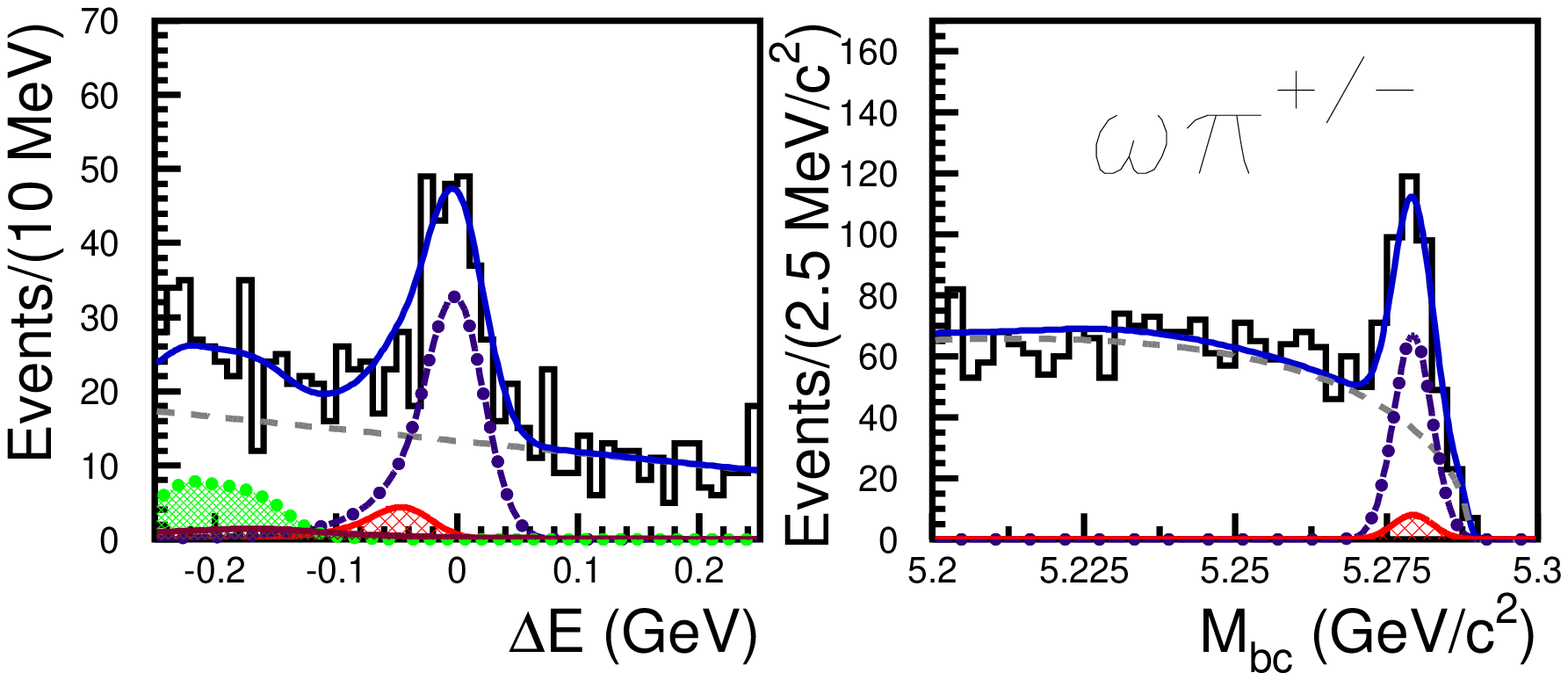}
\includegraphics[width=0.52\textwidth]{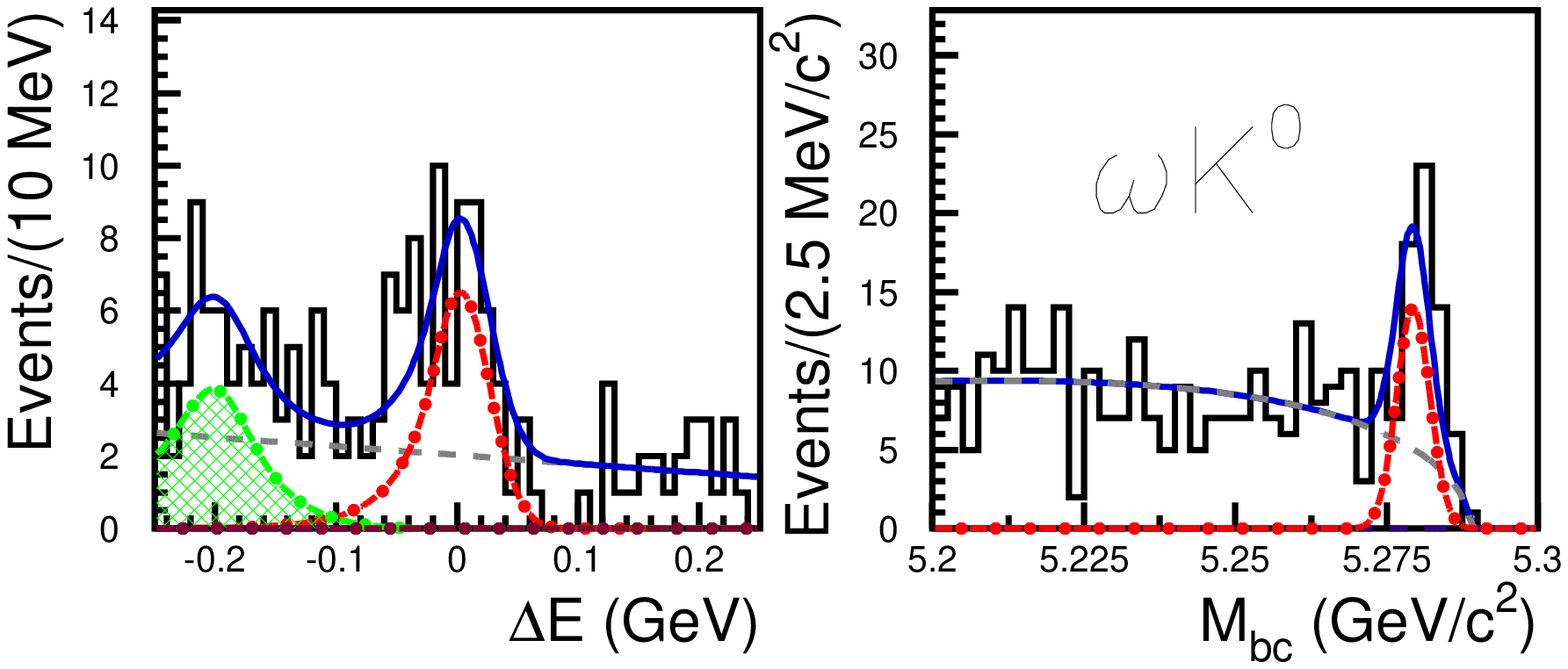}
\includegraphics[width=0.52\textwidth]{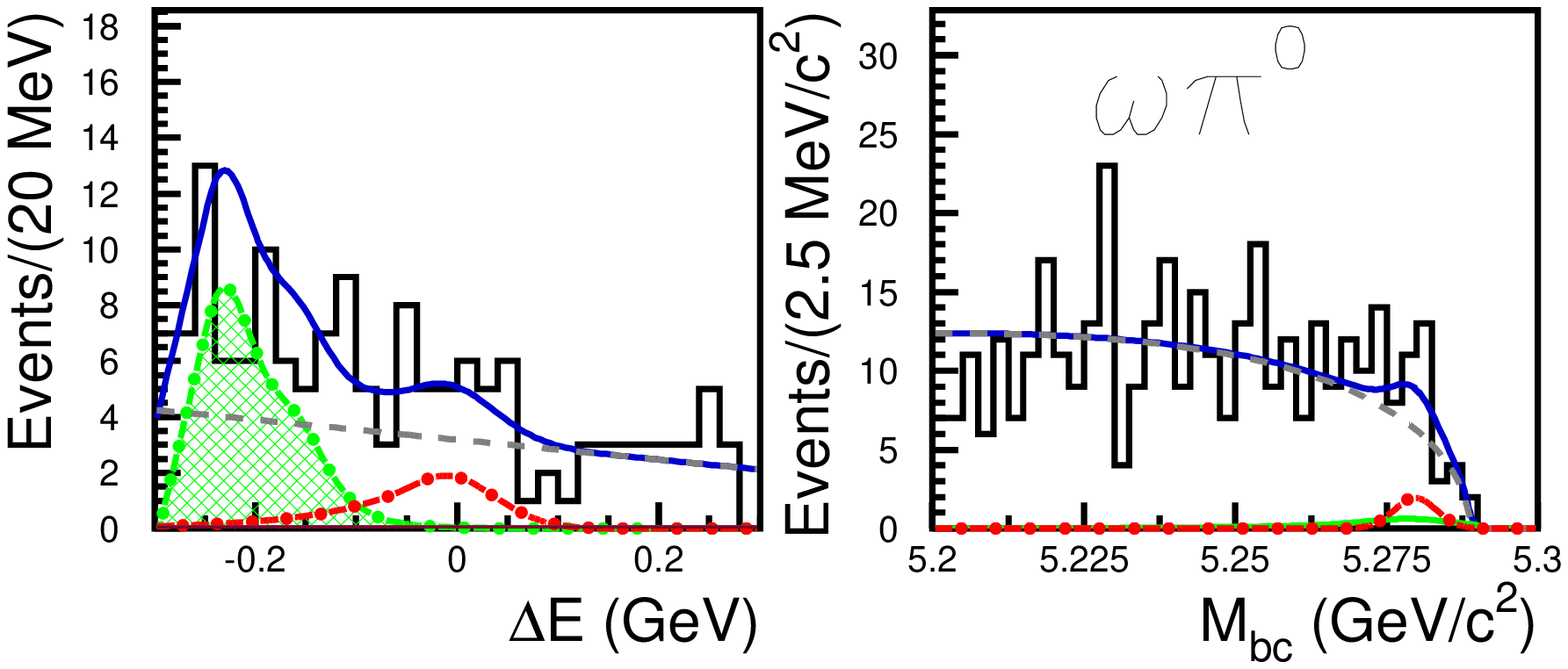}
\caption{Projections of fit results on $M_\mathrm{bc}$ (within the region of 
$|\Delta E|\leq0.1$, $-0.15\leq\Delta E\leq0.1$ for $\omega \pi^0$) and $\Delta E$ 
(within the region of $5.27<M_\mathrm{bc}<5.29$) for $\omega K^\pm$, $\omega \pi^\pm$, 
$\omega K^0$, and $\omega \pi^0$. Open histograms are data, solid curves are the fit functions, 
dash-dotted lines represent signals, the dashed lines show the $q\bar{q}$ continuum, light gray 
crossed-hatched areas represent rare $B$ decay backgrounds, dark gray hatched areas peaking
in the $M_{\mathrm bc}$ signal region represent reflections due to $K$-$\pi$ misidentification 
and the small dark-gray contributions in the negative $\Delta$E region of the charged modes are
due to generic $B\bar{B}$ background.} 
\label{omegafit}
\end{figure}
\begin{figure}
\includegraphics[width=0.52\textwidth]{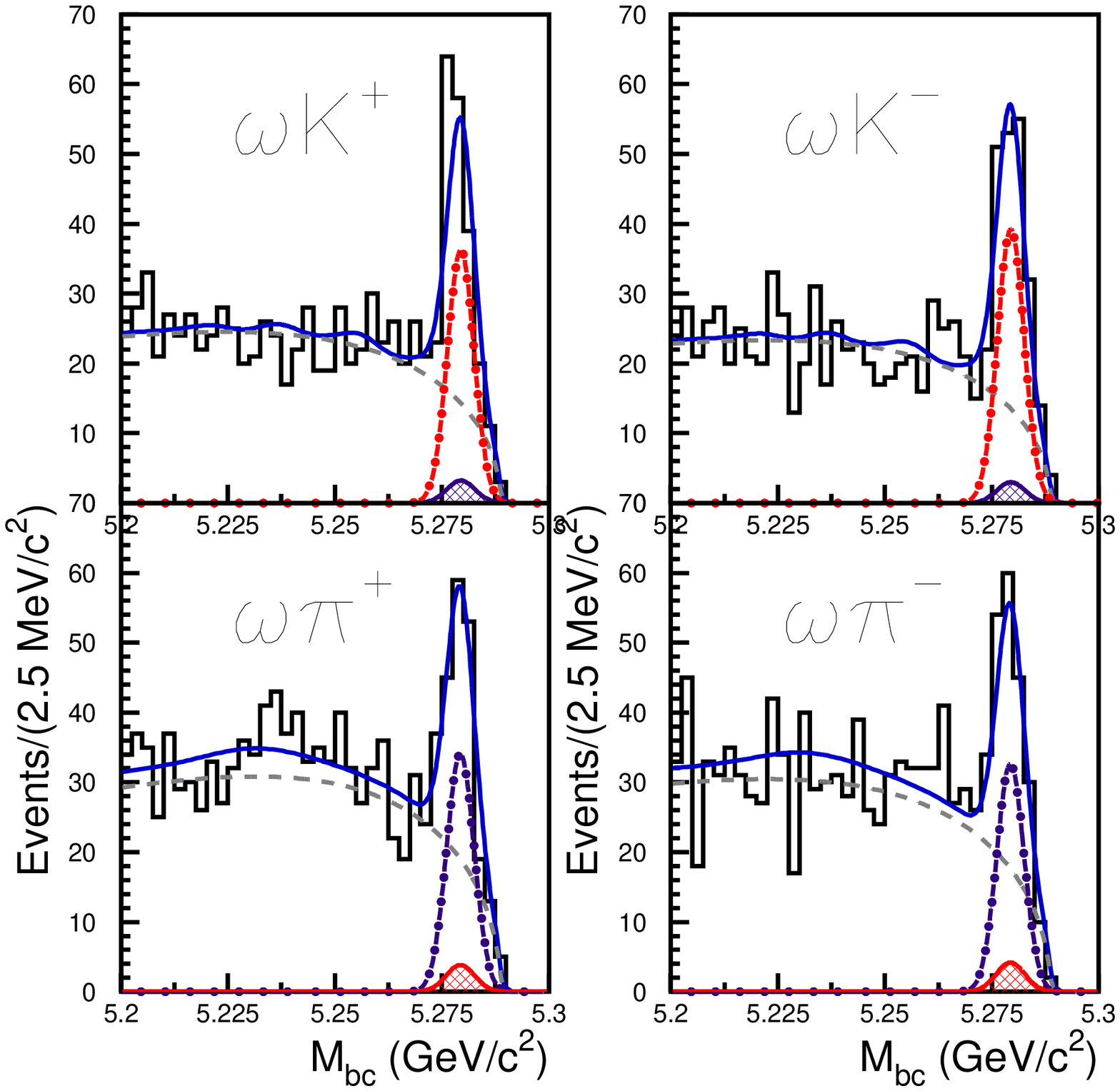}
\includegraphics[width=0.52\textwidth]{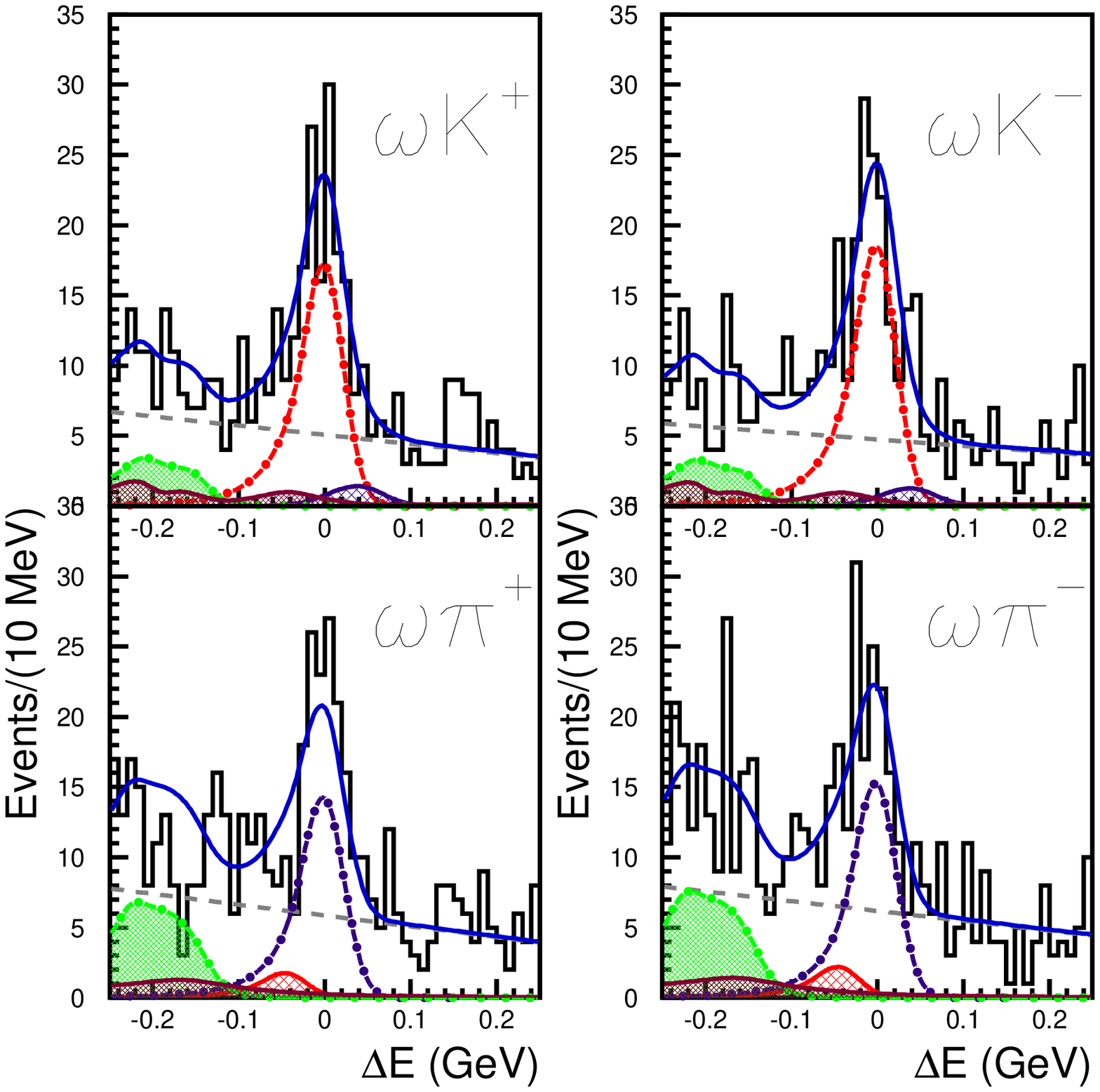}
\caption{Projections of fit results on $M_\mathrm{bc}$ (within the region of 
$|\Delta E|\leq0.1$, $-0.15\leq\Delta E\leq0.1$ for $\omega \pi^0$) and $\Delta E$ 
(within the region of $5.27<M_\mathrm{bc}<5.29$) for $\omega K^\pm$, $\omega \pi^\pm$, 
$\omega K^0$, and $\omega \pi^0$. Open histograms are data, solid curves are the fit functions, 
dash-dotted lines represent signals, the dashed lines show the $q\bar{q}$ continuum, light gray 
crossed-hatched areas represent rare $B$ decay backgrounds, dark gray hatched areas peaking
in the $M_{\mathrm bc}$ signal region represent reflections due to $K$-$\pi$ misidentification 
and the small dark-gray contributions in the negative $\Delta$E region are due to generic 
$B\bar{B}$ background.} 
\label{acpfit}
\end{figure}
We thank the KEKB group for the excellent operation of the
accelerator, the KEK cryogenics group for the efficient
operation of the solenoid, and the KEK computer group and
the National Institute of Informatics for valuable computing
and Super-SINET network support. We acknowledge support from
the Ministry of Education, Culture, Sports, Science, and
Technology of Japan and the Japan Society for the Promotion
of Science; the Australian Research Council and the
Australian Department of Education, Science and Training;
the National Science Foundation of China under contract
No.~10175071; the Department of Science and Technology of
India; the BK21 program of the Ministry of Education of
Korea, and the CHEP SRC program and Basic Reserch program 
(grant No. R01-2005-000-10089-0) of the Korea Science and
Engineering Foundation; the Polish State Committee for
Scientific Research under contract No.~2P03B 01324; the
Ministry of Science and Technology of the Russian
Federation; the Ministry of Higher Education, Science and Technology of the Republic of Slovenia;  
the Swiss National Science Foundation; the National Science Council and
the Ministry of Education of Taiwan; and the U.S.\ Department of Energy.
\end{document}

%% file: author-conf2005.tex
%%% Paper:    
%%% Journal:  Summer 2005 conference papers
%%% Contacts: 
%%% Non-responding authors or those who said NO are commented out.
%%% ====================================================================
%%% Click the RELOAD button on your web browser to see the updated file.
%%% ====================================================================
%%% Use \input{author} to insert this material into your latex file.
%%%%% Force institutions to appear in alphabetical order when typeset.
\affiliation{Aomori University, Aomori}
\affiliation{Budker Institute of Nuclear Physics, Novosibirsk}
\affiliation{Chiba University, Chiba}
\affiliation{Chonnam National University, Kwangju}
\affiliation{University of Cincinnati, Cincinnati, Ohio 45221}
\affiliation{University of Frankfurt, Frankfurt}
\affiliation{Gyeongsang National University, Chinju}
\affiliation{University of Hawaii, Honolulu, Hawaii 96822}
\affiliation{High Energy Accelerator Research Organization (KEK), Tsukuba}
\affiliation{Hiroshima Institute of Technology, Hiroshima}
\affiliation{Institute of High Energy Physics, Chinese Academy of Sciences, Beijing}
\affiliation{Institute of High Energy Physics, Vienna}
\affiliation{Institute for Theoretical and Experimental Physics, Moscow}
\affiliation{J. Stefan Institute, Ljubljana}
\affiliation{Kanagawa University, Yokohama}
\affiliation{Korea University, Seoul}
\affiliation{Kyoto University, Kyoto}
\affiliation{Kyungpook National University, Taegu}
\affiliation{Swiss Federal Institute of Technology of Lausanne, EPFL, Lausanne}
\affiliation{University of Ljubljana, Ljubljana}
\affiliation{University of Maribor, Maribor}
\affiliation{University of Melbourne, Victoria}
\affiliation{Nagoya University, Nagoya}
\affiliation{Nara Women's University, Nara}
\affiliation{National Central University, Chung-li}
\affiliation{National Kaohsiung Normal University, Kaohsiung}
\affiliation{National United University, Miao Li}
\affiliation{Department of Physics, National Taiwan University, Taipei}
\affiliation{H. Niewodniczanski Institute of Nuclear Physics, Krakow}
\affiliation{Nippon Dental University, Niigata}
\affiliation{Niigata University, Niigata}
\affiliation{Nova Gorica Polytechnic, Nova Gorica}
\affiliation{Osaka City University, Osaka}
\affiliation{Osaka University, Osaka}
\affiliation{Panjab University, Chandigarh}
\affiliation{Peking University, Beijing}
\affiliation{Princeton University, Princeton, New Jersey 08544}
\affiliation{RIKEN BNL Research Center, Upton, New York 11973}
\affiliation{Saga University, Saga}
\affiliation{University of Science and Technology of China, Hefei}
\affiliation{Seoul National University, Seoul}
\affiliation{Shinshu University, Nagano}
\affiliation{Sungkyunkwan University, Suwon}
\affiliation{University of Sydney, Sydney NSW}
\affiliation{Tata Institute of Fundamental Research, Bombay}
\affiliation{Toho University, Funabashi}
\affiliation{Tohoku Gakuin University, Tagajo}
\affiliation{Tohoku University, Sendai}
\affiliation{Department of Physics, University of Tokyo, Tokyo}
\affiliation{Tokyo Institute of Technology, Tokyo}
\affiliation{Tokyo Metropolitan University, Tokyo}
\affiliation{Tokyo University of Agriculture and Technology, Tokyo}
\affiliation{Toyama National College of Maritime Technology, Toyama}
\affiliation{University of Tsukuba, Tsukuba}
\affiliation{Utkal University, Bhubaneswer}
\affiliation{Virginia Polytechnic Institute and State University, Blacksburg, Virginia 24061}
\affiliation{Yonsei University, Seoul}
  \author{K.~Abe}\affiliation{High Energy Accelerator Research Organization (KEK), Tsukuba} % KEK
  \author{K.~Abe}\affiliation{Tohoku Gakuin University, Tagajo} % TohokuGakuin
  \author{I.~Adachi}\affiliation{High Energy Accelerator Research Organization (KEK), Tsukuba} % KEK
  \author{H.~Aihara}\affiliation{Department of Physics, University of Tokyo, Tokyo} % Tokyo
  \author{K.~Aoki}\affiliation{Nagoya University, Nagoya} % Nagoya
  \author{K.~Arinstein}\affiliation{Budker Institute of Nuclear Physics, Novosibirsk} % BINP
  \author{Y.~Asano}\affiliation{University of Tsukuba, Tsukuba} % Tsukuba
  \author{T.~Aso}\affiliation{Toyama National College of Maritime Technology, Toyama} % Toyama
  \author{V.~Aulchenko}\affiliation{Budker Institute of Nuclear Physics, Novosibirsk} % BINP
  \author{T.~Aushev}\affiliation{Institute for Theoretical and Experimental Physics, Moscow} % ITEP
% ...
% Lausanne
\collaboration{The Belle Collaboration}

%% file: omega_eps2005.bbl
\begin{thebibliography}{99}
%
%\bibitem{CKM}
%M. Kobayashi and T. Maskawa, Prog. Theor. Phys. 49, 652 (1973).
%
%\bibitem{SM}
%Y. Grossman, M.P. Worah, Phys. Lett. B 395, 241 (1997); D. London,
%A. Soni, Phys. Lett. B 407, 61 (1997).
\bibitem{QCDF} 
M. Beneke and M. Neubert, Nucl. Phys. B {\bf 675}, 333 (2003).

\bibitem{PQCD}  
C.-H. Chen, Phys. Lett. B {\bf 525}, 56 (2002);
Y.Y. Keum and A.I. Sanda, Phys. Rev.{\bf D67}, 054009 (2003).

\bibitem{chwang}
Belle Collaboration, C.H.~Wang {\it et al.},
Phys. Rev.{\bf D70}, 012001 (2004).

\bibitem{previous}BaBar Collaboration, B. Aubert {\it et al.}, Phys. 
Rev.{\bf D71}, 031103 (2005).

\bibitem{babar} BaBar Collaboration, B. Aubert {\it et al.}, 
hep-ex/0503018.

%\bibitem{btos} Belle Collaboration, K.-F. Chen  {\it et al}, Phys. Rev. 
%{\bf D72}, 012004 (2005).
%
\bibitem{KEKB} 
S.~Kurokawa and E.~Kikutani, Nucl. Instr. and. Meth. A {\bf 499}, 1 (2003),
and other papers included in this volume.
%
\bibitem{Belle} 
Belle Collaboration, A.~Abashian {\it et al.},
Nucl. Instr. and Meth. A {\bf 479}, 117 (2002).
%
\bibitem{thrust} We define the thrust axis for a collection of particles 
as the axis that maximizes the sum of the magnitude of the longitudinal
momenta with respect to the axis.
%  
\bibitem{SFW}
The Fox-Wolfram moments were introduced in
G.~C.~Fox and S.~Wolfram, Phys. Rev. Lett. {\bf 41}, 1581 (1978).
The Fisher discriminant used by Belle, based on modified Fox-Wolfram 
moments (SFW), is described in 
K.~Abe {\it et al.} (Belle Collab.), Phys. Rev. Lett. {\bf 87}, 101801 (2001) and
K.~Abe {\it et al.} (Belle Collab.), Phys. Lett. B {\bf 511}, 151 (2001). 
%
\bibitem{TaggingNIM}
Belle Collaboration, 
H. Kakuno {\it et al.}, Nucl. Instr. and Meth.A {\bf 533}, 516 (2004). 
%
\bibitem{crystal}
Crystal Ball Collaboration, J.E. Gaiser  {\it et al.},
Phys. Rev. D {\bf 34}, 711 (1986).
%
\bibitem{argus}ARGUS Collaboration, H. Albrecht {\it et al.}, Phys. Lett. B
 {\bf 241}, 278 (1990).
%
%\bibitem{PDG}
%S. Eidelman {\it et al.} (Particle Data Group), Phys. Lett. B {\bf 592, 1 (2004).
%
\end{thebibliography}
